\documentclass[a4paper]{article}

\usepackage{epsfig}
\usepackage{amssymb,amsmath,amsfonts}
\usepackage{multirow}
\usepackage{rotating}
\usepackage{graphicx}
\usepackage{lscape}
\usepackage{hyperref}
\usepackage{ulem}
\usepackage{xcolor}

\usepackage{cite} 

\usepackage{authblk} 

\usepackage{geometry}
\begin{document}
\title{\bf
Expected statistical uncertainties at future $e^+e^-$ colliders
}
\author[1]{Hieu Minh Tran%
	\thanks{hieu.tranminh@hust.edu.vn}}
\affil[1]{Hanoi University of Science and Technology, 
	1 Dai Co Viet Road, Hanoi, Vietnam
}
\author[2]{Sang Quang Dinh%
	}
\affil[2]{VNU University of Science, Vietnam National University - Hanoi,
	334 Nguyen Trai Road, Hanoi, Vietnam
}
\author[3]{Trang Quynh Trieu%
	}
\affil[3]{Nam Dinh College of Education, 
	813 Truong Chinh, Nam Dinh, Vietnam
}
\maketitle
\begin{abstract} 
In future colliders, the frontiers of luminosity and energy are extended
to explore the physics of elementary particles at extremely high precision,
and to discover new phenomena suggested from current experimental anomalies.
In this letter, we present a simple method to estimate the expected statistical uncertainties of scattering cross sections at future colliders using their conceptual design reports.
In particular, the expected statistical uncertainties of muon pair production cross section at the Future Circular Collider (FCC-ee) and the Circular Electron-Positron Collider (CEPC) are calculated.
The results can be used 
to set a goal for systematic uncertainty improvement,
to determine the standard model parameters accurately, and
to identify the viable parameter space of new physics models.

\end{abstract}
\newpage
%
%
%

\textit{Introduction $-$}
%
%
%
The particle colliders such as the Large Electron-Positron (LEP) Collider, the Tevatron, and the current Large Hadron Collider (LHC) have been playing important roles in the development of our understanding of the most fundamental particles and their interactions.
In order to determine the properties of the standard model (SM) particles more precisely, as well as to probe possible novel physical effects, or providing stringent constraints on new physics models,
new generations of colliders with higher center of mass energies and luminosities than the current frontiers are necessary.
Several projects are proposed including 
the Future Circular Collider (FCC) \cite{FCC:2018byv,FCC:2018evy}
and the Circular Electron Position Collider (CEPC)
\cite{CEPCStudyGroup:2018rmc,CEPCStudyGroup:2023quu}
to achieve that goal.
In their conceptual design reports, the important information about tentative energy scales and corresponding luminosities was provided.
To investigate the sensitivity of a future collider for a given physical process, it is necessary to know not only the relevant observables but also its uncertainties.
While the systematic uncertainty depends on detector technologies and measurement methods, the statistical uncertainty evaluated based on the expected number of signal events is unavoidable.
Since the statistical uncertainty is reduced by increasing either the number of independent experiments or the integrated luminosity of colliders, higher luminosity is one of the target in the development of future colliders.
Predictions of these uncertainties at future colliders can be obtained using complicated simulation tools \cite{Campbell:2022qmc}.

In this letter, 
we present a simple method to evaluate the statistical uncertainty at future colliders using their technical design reports and the known result of similar older experiments.
This is an alternative approach to the one using a combination of event generators and detector simulations.
The method is verified against existing experimental data at the LEP collider.
%
%
It is then applied to estimate the statistical uncertainty of the cross section of the muon pair production at the FCC-ee and the CEPC.

%
%

\begin{figure}[h!]
	\begin{center}
		\includegraphics[scale=0.7]{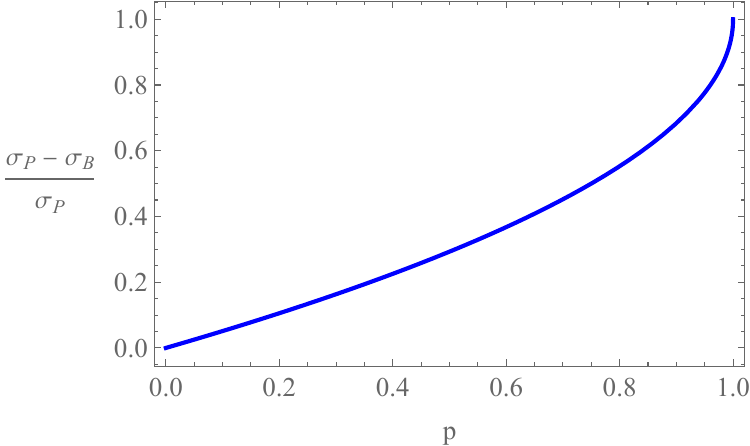}
	\end{center}
	\caption{The relative difference between the standard deviations of the Poisson and the binomial distributions as a function of the probability $p$.}
	\label{sd}
\end{figure}

\textit{Statistics of Bernoulli trials $-$}
Let us call $n$ the number of independent experiments, of which the results are Bernoulli trials with two possible outcomes (success, or failure).
Assuming the probability for one experiment leading to a success is $p$, the probability that $n$ experiments results in $k$ successes and $n-k$ failures is given by the binomial distribution:
\begin{align}
f_B (k;n,p) = 
	\frac{n!}{k! (n-k)!}
	p^k (1-p)^{n-k} .
	\label{binomial}
\end{align}
The mean value and the standard deviation of the binomial distribution are respectively
\begin{align}
	N = n p, \qquad
	\sigma_B = \sqrt{n p (1-p)}.
\end{align}
According to the Poisson limit theorem, when $n$ tends to infinity and $N$ is kept constant (implying $p$ is approaching zero), the binomial distribution converges to the Poisson distribution:
\begin{align}
	f_B (k;n,p) 
	\xrightarrow[N = \text{const}]{n \rightarrow \infty}
	f_P (k;N) =
	e^{-N} \frac{N^k}{k!} .
	\label{poisson}
\end{align}
The relative difference between standard deviations of the two distributions,
$\frac{\sigma_P - \sigma_B}{\sigma_P}$ is independent of $n$.
It is shown in Figure \ref{sd} as a function of the probability $p$.
The difference between the standard deviations of two distributions is
\begin{align}
	\sigma_P - \sigma_B 
	=
	\sqrt{N} - \sqrt{np(1-p)}
	\approx
	\frac{p}{2} \sigma_P .
\end{align}
This explains why in the low region of $p$ in Figure \ref{sd} the ratio appears roughly as a linear function of $p$.
%
In the limit of vanishing $p$, the two standard deviations become the same.
Therefore, when $p$ is small enough, we can approximate the standard deviation of the binomial distribution by that of the Poisson distribution:
\begin{align}
	\sigma_B \approx \sigma_P = \sqrt{N} .
	\label{standard-deviation}
\end{align}

In practice at colliders, $n$ is regarded as the number of collisions between incoming particles.
Considering a certain final state $X$, the outcome of each collision can be classified as either $X$ or non-$X$.
From this point of view, each particle collision is one Bernoulli trial.
After $n$ collisions, the number of $X$-events recorded by detectors is $k$.
Obviously, $k$ is a random number obeying the binomial distribution (\ref{binomial}).
The recorded value of $k$ is used to evaluate its (expected) mean value:
\begin{align}
	N \approx k.
\end{align}
Since $n$ is very large for an ordinary analysis at particle colliders, we can approximate the binomial distribution by the Poisson distribution (\ref{poisson})
if the probability of obtaining the final state $X$ is very small.
The statistical uncertainty of the number of $X$-events is given by the standard deviation of the Poisson distribution, namely
\begin{align}
	\Delta N \approx \sigma_P = \sqrt{N}.
\end{align}

%
%
%






\textit{Statistical uncertainty of cross section $-$}
The cross section of a collision producing the final state $X$ is determined by the number of observed signal events ($N$) and the integrated luminosity ($L$) as
\begin{align}
	\sigma_X = \frac{N}{\epsilon L} ,
\end{align}
where $\epsilon$ is the overall efficiency encoding the information of detector acceptance and selection efficiency.
Assuming negligible uncertainties of $L$ and $\epsilon$,
the statistical uncertainty of this cross section is the direct consequence of the statistical uncertainty of $N$,
namely
\begin{align}
	\Delta \sigma_X \approx \frac{\Delta N}{\epsilon L} 
	\approx
	\sqrt{\frac{\sigma_X}{\epsilon L}} .
	\label{approx_err}
\end{align}
Applying Eq. (\ref{approx_err}) to two measurements, we arrive at the following relationship
\begin{align}
	\frac{\Delta \sigma_1}{\Delta \sigma_0} = 
	\sqrt{
		\frac{\sigma_1/(\epsilon_1 L_1)}
		{\sigma_0/(\epsilon_0 L_0)}} .
	\label{D-ratio}
\end{align}
In general, the subscripts ``0'' and ``1'' in Eq. (\ref{D-ratio}) denote any kind of differences between the two measurements.
There are remarks that we would like to emphasize regarding the above equation:
($i$) the uncertainty is the statistical only, which is related to the number of signal events,
($ii$) the condition of the Poisson approximation must be satisfied, i.e. $N \ll n$, implying that the cross sections $\sigma_{0,1}$ must be small compared to the total inclusive cross section.
%
%

In some realistic situation where either the efficiency $\epsilon_0$ or $\epsilon_1$ is unknown, such as the case of future colliders,
the usage of Eq. (\ref{D-ratio}) is restricted.
It is useful to assume for simplicity that the overall efficiencies of the two experiments are approximately the same
\begin{align}
	\epsilon_0 \approx \epsilon_1.
	\label{eff_assumption}
\end{align}
This assumption is reasonable if we consider the same physical process relevant to $\sigma_0$ and $\sigma_1$, while the energy scales and the luminosities of these two experiments vary.
For example, let us consider the case where
the result of one experiment is expected in the future, while the other was already known.
By the virtue of Eq. (\ref{D-ratio}), the known experimental result is the reference point that can be used to determined the statistical uncertainty of the future experiment.
Substituting the assumption (\ref{eff_assumption}) into Eq. (\ref{D-ratio}),
the estimated statistical uncertainty of the cross section $\sigma$ of a scattering process in a future measurement with integrated luminosity $L$, is found to be
\begin{align}
	\Delta \sigma_\text{est} = 
	\Delta \sigma_0
	\sqrt{\frac{\sigma /L }{\sigma_0 / L_0}} .
	\label{est_uncer}
\end{align}
Here, $\sigma_0$, $\Delta\sigma_0$ and $L_0$ are the reference data of the cross section, the statistical uncertainty and the  luminosity of a known experiment.
The cross section $\sigma$ is calculated theoretically, and the luminosity $L$ is given in the conceptual design report of the relevant future collider.



\begin{figure}[h!]
	\begin{center}
		\includegraphics[scale=0.8]{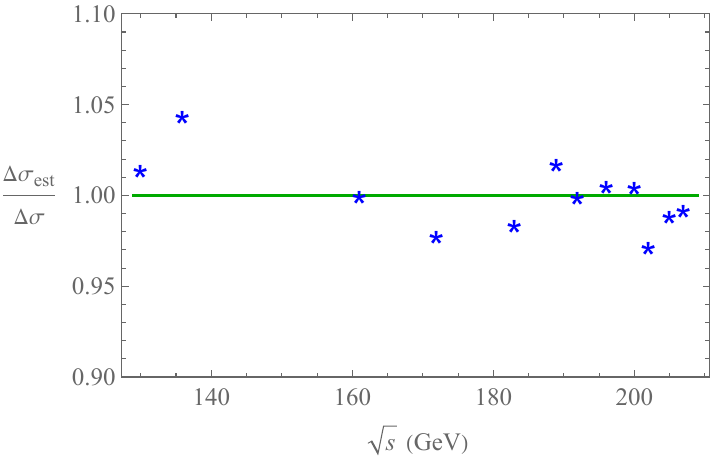}
	\end{center}
	\caption{The ratio between the estimated statistical uncertainty of $\sigma(e^+e^-\rightarrow\mu^+\mu^-)$ and the recorded one at the DELPHI experiment \cite{DELPHI:2005wxt}.}
	\label{ratio}
\end{figure}

To demonstrate the validity of the method, we consider the muon pair production process, 
$e^+ e^- \rightarrow \mu^+ \mu^-$, measured by the DELPHI collaboration 
\cite{DELPHI:2005wxt}
at the LEP collider.
Due to initial state radiations, the effective colliding energy becomes smaller than the nominal one.
In this case, the cut on the effective colliding energy,
$\sqrt{s'}>75$ GeV, was applied by the DELPHI collaboration.
The data at the center-of-mass energy $\sqrt{s_0} = 192$ GeV is taken to be the reference point with
the measured cross section $\sigma_0(e^+e^-\rightarrow\mu^+\mu^-) = 7.37$ pb,
the statistical uncertainty $\Delta \sigma_0 = 0.61$ pb, and
the luminosity
$L_0 = 25.79$ pb$^{-1}$.
In Table \ref{estimateDELPHI}, 
the first two columns show the center-of-mass energies ($\sqrt{s}$) and the recorded statistical uncertainties of the cross section ($\Delta\sigma$) at the DELPHI experiment \cite{DELPHI:2005wxt}.
The estimated statistical uncertainties ($\Delta\sigma_\text{est}$) 
calculated using Eq. (\ref{est_uncer})
are presented in the third column.
By comparing the recorded and the estimated uncertainties in the last two columns, we see that they are in good agreement.
\begin{table}[h!]
\begin{center}
	\begin{tabular}{ ccc } 
	$\sqrt{s}$ (GeV) & $\Delta\sigma$ (pb)& $\Delta\sigma_\text{est}$ (pb) \\ 
	\hline 
	130	&	3.2 	&	3.248	\\
	136	&	2.6 	&	2.716	\\
	161	&	1.1 	&	1.100	\\
	172 &	1.1 	&	1.077	\\
	183	&	0.46	&	0.453	\\
	189	&	0.24	&	0.244	\\
	192	&	0.61	&	0.610	\\
	196	&	0.32	&	0.322	\\
	200	&	0.33	&	0.332	\\
	202	&	0.44	&	0.428	\\
	205	&	0.31	&	0.307	\\
	207	&	0.23	&	0.228	
	\end{tabular}
	\caption{The recorded and the estimated statistical uncertainties at the DELPHI experiment 
	\cite{DELPHI:2005wxt}
	with the acceptance $\sqrt{s'}>75$ GeV.}
	\label{estimateDELPHI}
\end{center}
\end{table}
In Figure \ref{ratio}, the ratios between estimated statistical uncertainties and recorded ones,
$\frac{\Delta\sigma_\text{est}}{\Delta\sigma}$,
at the DELPHI experiment are shown for various center-of-mass energies.
Since all of these ratios are close to unity, the method works well in this case  with the relative deviations being less than 5\%.

\textit{Statistical uncertainties at FCC-ee and CEPC $-$}
For the case of future circular colliders \cite{Shiltsev:2019rfl}, such as the FCC-ee
\cite{FCC:2018evy} and 
the CEPC 
\cite{CEPCPhysicsStudyGroup:2022uwl}, 
the advantage of high luminosity helps to significantly reduce statistical uncertainties.
Since the efficiencies at these collider have not been known yet, 
the condition in Eq. (\ref{eff_assumption}) is assumed for simplicity.
This requires us to consider the same  physical process for the two experiments in the estimation method.
Here, we focus on the scattering process $e^+e^- \rightarrow \mu^+\mu^-$ at the FCC-ee and the CEPC, 
and predict the statistical uncertainty of its cross section.

To apply Eq. (\ref{est_uncer}) for the determination of statistical uncertainties at the future colliders, the reference point from the LEP 2 combined result \cite{ALEPH:2013dgf} is taken.
Specifically, we choose
the reference point corresponding to the center-of-mass energy $\sqrt{s_0}=172$ GeV, 
the luminosity 
$L_0 = 10 \text{ pb}^{-1}$, and
the measured cross section 
$\sigma_0(e^+e^- \rightarrow \mu^+\mu^-) = 3.562$ pb together with its statistical uncertainty
$\Delta \sigma_0 = 0.331$ pb. 
We employ the ZFITTER package version 6.42
\cite{Bardin:1992jc, Bardin:1999yd, Arbuzov:2005ma, Akhundov:2013ons} 
to calculate the SM prediction of the cross section $\sigma(e^+e^- \rightarrow \mu^+\mu^-)$ at various colliding energies according to the conceptual design reports 
\cite{FCC:2018evy,CEPCStudyGroup:2023quu}
of the FCC-ee and the CEPC.
In this calculation, the quantum loop corrections and the effects of initial state radiations are taken into account.
%
%
%
In our analyses, due to the assumption  (\ref{eff_assumption}), the same cut on the effective colliding energy as that in the LEP 2 experiment
is considered, 
namely $\sqrt{s'/s} > 0.85$ 
\cite{ALEPH:2013dgf}.

In the first three columns of Tables \ref{FCC-ee} and \ref{CEPC}, the center-of-mass energies and the luminosities corresponding to four running modes ($Z$, $W^+W^-$, $ZH$, and $t\bar{t}$) of the FCC-ee and the CEPC colliders are given according to their technical design reports 
\cite{FCC:2018byv, CEPCStudyGroup:2023quu}.
The SM prediction of cross section for the scattering process $e^+e^-\rightarrow \mu^+\mu^-$ are presented in the fourth columns.
The estimated statistical uncertainties $\Delta\sigma_\text{est}$ of the cross sections are shown in the fifth column.
By comparing with the results at the LEP collider around the $W^+W^-$ threshold \cite{ALEPH:2013dgf}, 
we observe that the statistical uncertainties at the FCC-ee and CEPC colliders are about  $\mathcal{O}(10^3)$ times smaller than those at the LEP collider.
Additionally, in the last columns of these two tables, the relative statistical uncertainties, 
$\delta_\text{est} = \frac{\Delta\sigma_\text{est}}{\sigma}$, 
are provided, showing that the cross section can be measured at these two future colliders precisely at an unprecedented level.

\begin{table}[h]
	\begin{center}
		\begin{tabular}{c|ccccc}
			Mode	&	$\sqrt{s}$ (GeV)	& $L$ (ab$^{-1}$)	& $\sigma$ (pb)	&
			$\Delta \sigma_\text{est}$ (pb)	& $\delta_\text{est}$	(\textperthousand) \\
			\hline
			\multirow{3}{*}{$Z$}	&	87.7	&	40	&	175.70268	&	0.00116	&	0.0066	\\
			&	91.2	&	80	&	1477.65318	&	0.00238	&	0.0016	\\
			&	93.9	&	40	&	451.36645	&	0.00186	&	0.0041	\\
			\hline	
			\multirow{2}{*}{$W^+W^-$}	&	157.5	&	6	&	4.94381	&	0.00050	&	0.10	\\
			&	162.5	&	6	&	4.60363	&	0.00049	&	0.11	\\
			\hline	
			$ZH$	&	240	&	5	&	1.89873	&	0.00034	&	0.18	\\
			\hline
			\multirow{9}{*}{$t\bar{t}$}	&	340	&	0.025	&	0.91479	&	0.00335	&	3.67	\\
			&	341	&	0.025	&	0.90917	&	0.00334	&	3.68	\\
			&	341.5	&	0.025	&	0.90637	&	0.00334	&	3.68	\\
			&	342	&	0.025	&	0.90358	&	0.00333	&	3.69	\\
			&	343	&	0.025	&	0.89802	&	0.00332	&	3.70	\\
			&	343.5	&	0.025	&	0.89525	&	0.00332	&	3.71	\\
			&	344	&	0.025	&	0.89247	&	0.00331	&	3.71	\\
			&	345	&	0.025	&	0.88684	&	0.00330	&	3.72	\\
			&	365	&	1.5	&	0.78971	&	0.00040	&	0.51
		\end{tabular}
		\caption{Cross section and statistical uncertainty for the process
			$e^+e^- \rightarrow \mu^+\mu^-$ at the FCC-ee with the acceptance $\sqrt{s'/s}>0.85$.}
		\label{FCC-ee}
	\end{center}
\end{table}

\begin{table}[h!]
	\begin{center}
		\begin{tabular}{c|ccccc}
			Mode	&	$\sqrt{s}$ (GeV)	& $L$ (ab$^{-1}$)	& $\sigma$ (pb)	&	$\Delta \sigma_\text{est}$ (pb)	& $\delta_\text{est}$ (\textperthousand)	\\
			\hline
			$Z$	&	91.2	&	100	&	1477.65318	&	0.00213	&	0.0014	\\
			$W^+W^-$	&	160	&	6.9	&	4.77171	&	0.00046	&	0.097	\\
			$ZH$	&	240	&	21.6	&	1.89873	&0.00016	&	0.087	\\
			$t\bar{t}$	&	360	&	1	&	0.81234	&	0.00050	&	0.62
		\end{tabular}
		\caption{Cross section and statistical uncertainty for the process
			$e^+e^- \rightarrow \mu^+\mu^-$ at the CEPC with the acceptance $\sqrt{s'/s}>0.85$.}
		\label{CEPC}
	\end{center}
\end{table}

Such small statistical uncertainties at the FCC-ee and the CEPC would help to accurately measure the SM parameters, 
enabling the possibility to unveil  hidden imprints of new physics, and constraining models beyond the SM.
In an optimized experiment, the systematic uncertainty should be smaller or equivalent to the statistical one.
The estimated statistical uncertainties are useful to identify the goal to reduce the systematic uncertainties in the future.
Therefore, beside the phenomenological usefulness,
the method and the result presented here are also important for the detector research and development.

\textit{Conclusions $-$}
Statistical uncertainty, an inescapable type of error in experiments, 
can be reduced significantly in the future with new generations of circular colliders by increasing the luminosity up to the level of $\mathcal{O}$(ab$^{-1}$) or higher.
In this letter,
we have presented a simple method to evaluate the statistical uncertainty of scattering cross section at future colliders by using measured data at previous colliders.
The method is verified by data from the DELPHI experiment at the LEP collider.
We have calculated the expected statistical uncertainty of the 
$e^+e^- \rightarrow \mu^+\mu^-$ process at the FCC-ee and the CEPC using their conceptual design reports.
The results can be used to determine the sensitivity of these future colliders in 
precisely testing the SM, 
and constraining new physics models.
They also set the criterion to improve the systematic uncertainty at these future colliders.

\section*{Acknowledgements}
This research is funded by Vietnam 
National Foundation for Science and Technology Development (NAFOSTED)
under grant number 103.01-2023.75.

\bibliographystyle{JHEP} 
\bibliography{references}

\providecommand{\href}[2]{#2}\begingroup\raggedright\begin{thebibliography}{10}

\bibitem{FCC:2018byv}
{\scshape FCC} collaboration,  \textit{{FCC Physics Opportunities}: {Future
  Circular Collider Conceptual Design Report Volume 1}},
  \href{https://doi.org/10.1140/epjc/s10052-019-6904-3}{Eur. Phys. J. C
  {\bfseries 79} (2019) 474}.

\bibitem{FCC:2018evy}
{\scshape FCC} collaboration,  \textit{{FCC-ee: The Lepton Collider}: {Future
  Circular Collider Conceptual Design Report Volume 2}},
  \href{https://doi.org/10.1140/epjst/e2019-900045-4}{Eur. Phys. J. ST
  {\bfseries 228} (2019) 261}.

\bibitem{CEPCStudyGroup:2018rmc}
{\scshape CEPC Study Group} collaboration,  \textit{{CEPC Conceptual Design
  Report: Volume 1 - Accelerator}},
  \href{https://arxiv.org/abs/1809.00285}{{\ttfamily 1809.00285}}.

\bibitem{CEPCStudyGroup:2023quu}
{\scshape CEPC Study Group} collaboration,  \textit{{CEPC Technical Design
  Report: Accelerator}},
  \href{https://doi.org/10.1007/s41605-024-00463-y}{Radiat. Detect. Technol.
  Methods {\bfseries 8} (2024) 1}
  [\href{https://arxiv.org/abs/2312.14363}{{\ttfamily 2312.14363}}].

\bibitem{Campbell:2022qmc}
J.M.~Campbell et~al.,  \textit{{Event generators for high-energy physics
  experiments}}, \href{https://doi.org/10.21468/SciPostPhys.16.5.130}{SciPost
  Phys. {\bfseries 16} (2024) 130}
  [\href{https://arxiv.org/abs/2203.11110}{{\ttfamily 2203.11110}}].

\bibitem{DELPHI:2005wxt}
{\scshape DELPHI} collaboration,  \textit{{Measurement and interpretation of
  fermion-pair production at LEP energies above the Z resonance}},
  \href{https://doi.org/10.1140/epjc/s2005-02461-0}{Eur. Phys. J. C {\bfseries
  45} (2006) 589} [\href{https://arxiv.org/abs/hep-ex/0512012}{{\ttfamily
  hep-ex/0512012}}].

\bibitem{Shiltsev:2019rfl}
V.~Shiltsev and F.~Zimmermann,  \textit{{Modern and Future Colliders}},
  \href{https://doi.org/10.1103/RevModPhys.93.015006}{Rev. Mod. Phys.
  {\bfseries 93} (2021) 015006}
  [\href{https://arxiv.org/abs/2003.09084}{{\ttfamily 2003.09084}}].

\bibitem{CEPCPhysicsStudyGroup:2022uwl}
{\scshape CEPC Physics Study Group} collaboration,  \textit{{The Physics
  potential of the CEPC. Prepared for the US Snowmass Community Planning
  Exercise (Snowmass 2021)}},  in \emph{{Snowmass 2021}}, 5, 2022
  [\href{https://arxiv.org/abs/2205.08553}{{\ttfamily 2205.08553}}].

\bibitem{ALEPH:2013dgf}
{\scshape ALEPH, DELPHI, L3, OPAL, LEP Electroweak} collaboration,
  \textit{{Electroweak Measurements in Electron-Positron Collisions at
  W-Boson-Pair Energies at LEP}},
  \href{https://doi.org/10.1016/j.physrep.2013.07.004}{Phys. Rept. {\bfseries
  532} (2013) 119} [\href{https://arxiv.org/abs/1302.3415}{{\ttfamily
  1302.3415}}].

\bibitem{Bardin:1992jc}
D.Y.~Bardin et~al.,  \textit{{ZFITTER: An Analytical program for fermion pair
  production in e+ e- annihilation}},
  \href{https://arxiv.org/abs/hep-ph/9412201}{{\ttfamily hep-ph/9412201}}.

\bibitem{Bardin:1999yd}
D.Y.~Bardin, P.~Christova, M.~Jack, L.~Kalinovskaya, A.~Olchevski, S.~Riemann
  et~al.,  \textit{{ZFITTER v.6.21: A Semianalytical program for fermion pair
  production in $e^+ e^-$ annihilation}},
  \href{https://doi.org/10.1016/S0010-4655(00)00152-1}{Comput. Phys. Commun.
  {\bfseries 133} (2001) 229}
  [\href{https://arxiv.org/abs/hep-ph/9908433}{{\ttfamily hep-ph/9908433}}].

\bibitem{Arbuzov:2005ma}
A.B.~Arbuzov, M.~Awramik, M.~Czakon, A.~Freitas, M.W.~Grunewald, K.~Monig
  et~al.,  \textit{{ZFITTER: A Semi-analytical program for fermion pair
  production in e+ e- annihilation, from version 6.21 to version 6.42}},
  \href{https://doi.org/10.1016/j.cpc.2005.12.009}{Comput. Phys. Commun.
  {\bfseries 174} (2006) 728}
  [\href{https://arxiv.org/abs/hep-ph/0507146}{{\ttfamily hep-ph/0507146}}].

\bibitem{Akhundov:2013ons}
A.~Akhundov, A.~Arbuzov, S.~Riemann and T.~Riemann,  \textit{{The ZFITTER
  project}}, \href{https://doi.org/10.1134/S1063779614030022}{Phys. Part. Nucl.
  {\bfseries 45} (2014) 529} [\href{https://arxiv.org/abs/1302.1395}{{\ttfamily
  1302.1395}}].

\end{thebibliography}\endgroup
\end{document}